**Effect of W in Cu-Zr-W thin films: Molecular dynamics simulations and experimental verification**

Hassan Ataalite,[*] Jiri Houska, Deepika Thakur, Michaela Cervena and Petr Zeman

Department of Physics and NTIS - European Centre of Excellence
University of West Bohemia in Pilsen, Univerzitni 8, 301 00 Pilsen, Czech Republic

**ABSTRACT**. We investigate the effects of W incorporation into Cu-Zr thin film metallic glasses using molecular dynamics (MD) simulations combined with magnetron sputtering. All studies are carried out in the whole range of W concentrations (0 to 100 at. %) and the MD studies also in a wide range of incident energies (1 to 500 eV) and deposition angles (0 to 60°). Calculated X-ray diffractograms, packing factor, short-range order (bonding fractions and coordination numbers), medium-range order (network ring and common neighbor statistics) and stress are correlated with measured X-ray diffractograms and properties (hardness, hardness/Young's modulus ratio and elastic recovery). The simulations explain the experimental results at the atomic level and provide a lot of information that is not available experimentally. Special attention is paid to non-monotonic dependencies on the elemental composition and incident energy. Collectively, the results explain the role of W in modifying the structure and improving the mechanical performance of Cu-Zr metallic glasses, predict optimum compositions which maximize some of the mechanical properties, and contribute to the development of advanced materials for various applications.

## I. INTRODUCTION.

Metallic glasses (MGs) are metal-based materials engineered to exhibit a combination of metallic and glass-like characteristics, resulting in an amorphous structure. This amorphous nature arises through the rapid cooling, or melt-quenching, of metal alloys, a process that has been key to the discovery and development of MGs since the 1960s [1]. A key factor in attracting global interest in MGs is their ability to combine several properties, such as high elastic limit, better mechanical properties, and greater corrosion resistance than their crystalline counterparts [2,3]. Additional factors include self-healing (healing small cracks or defects) behavior in the softened state above glass transition temperature, higher electrical conductivity, improved plasticity and fatigue resistance of thin film MGs than their bulk counterparts [4–6]. Recently, Cu-Zr metallic glasses have been a good example of this material class, attracting considerable attention due to their unique properties. They can be prepared in the bulk form, e.g. by melt-quenching [7] or in the thin film form, e.g. by magnetron sputtering [8]. Extensive research and experimentation have been carried out to improve the properties of Cu-Zr metallic glasses by incorporating other additional alloying elements such as Al, Fe, Ti, and Co [9,10]. Incorporating W is particularly important: it has the highest melting point among metals (3422 °C) [11] due to a strong covalent component of the covalent-metallic bonding, which constitutes a basis for its specific role in improving the thermal stability of Cu-Zr metallic glasses [12]. This property enables the Cu-Zr-W alloys to resist crystallization at high temperatures more effectively than other materials.

Generally, MGs are characterized by a disordered structure which has local atomic order without grains and their boundaries and does not exhibit any long-range crystalline order [13]. Thus, the properties of MGs are to a large extent defined by the existence of icosahedral clusters, which cease the formation of crystalline structure. Previously, the icosahedral clusters have been observed experimentally (e.g. X-ray/neutron scattering [14]) and through computer simulations (e.g. ab initio [15] and empirical interatomic interactions [16]). In addition, during the cooling process, these clusters get connected to each other by sharing their vertex, edge, intercross, or face to form a chain of an icosahedral network [17], ultimately resulting in the medium-range order.

Computer simulations provide a good understanding of the formation of metallic glass thin films. They allow a detailed description of the film structure at the atomic scale, which is often inaccessible experimentally. Classical Molecular Dynamics (MD) based on empirical interatomic potentials is a powerful method to directly reproduce the atom-by-atom thin film growth processes (not only to indirectly predict the film structure by other simulation algorithms such as Monte Carlo [18] or melt-quench [19]). Many research papers demonstrate the successful use of MD simulation for different categories of materials such as covalent: a-C [20], Si [21], SiC [22] or SiNH [23], ionic: α-$Al_2O_3$ [24], $ZrO_2$ [25,26], $TiO_2$ [27], MgO [28] or ZnO [29,30] and metallic: Cu-Zr [16,31] or Cu-Zr-Al [9,32]. However, despite this high potential, despite the aforementioned specific role of W, and despite the availability of empirical experimental results (see below), growth simulations of Cu-Zr-W are not yet available.

The present paper combines MD simulations of the atom-by-atom growth of Cu-Zr-W thin films with experimental results of their preparation by magnetron sputtering. We investigate the effect of W content, [W], in Cu-Zr-W thin film MGs (low [W]) and crystalline films (high [W]) on the structural and mechanical characteristics. The studies are carried out by varying not only [W] but also energy and deposition angle of arriving atoms. Calculated X-ray diffractograms, packing factor, short-range order (bonding fractions and coordination numbers), medium-range order (network ring and common neighbor statistics) and stress are correlated with measured diffractograms and properties (hardness, hardness/Young's modulus ratio and elastic recovery). Special attention is paid to non-monotonic dependencies on the aforementioned parameters.

*Contact author: hataali@ntis.zcu.cz

## II. METHODOLOGY

### A. Modelling

In this work, MD simulations were performed for modeling the atom-by-atom growth of Cu-Zr-W thin films using the LAMMPS (Large scale Atomic/Molecular Massively Parallel Simulator) package [33] and the snapshots were prepared using OVITO open-source tool [34]. The embedded atom method (EAM) [35] is used to describe the interaction energy between atoms. We employed the parameterization of the EAM potential developed by Zhou et al. [36]. This potential is realistic in reproducing the structural and mechanical properties of Cu-Zr metallic glasses thin films [37,38], W-Cu solid solution [39,40] and Mo-Ta-Ti-W-Zr refractory high-entropy alloy (RHEA) [41], at a good agreement with available experimental data. Furthermore, the correct prediction of preferred crystal structures, lattice constants (within 1%) and energies (within 2%) of Cu, Zr, and W elements by this potential has been also crosschecked (see Supplemental Materials [42] including Refs. [43–45]).

The ratio of [Cu]/[Zr] is approximately equal to 1, meaning that the composition formula became $Cu_xZr_xW_{100-2x}$, where $x$ = 0-50. Two growth templates are used in all simulations, leading to similar results at low [W] but intentionally different results at high [W]. The first and more important template of 1920 atoms of c-W(011) allowed us to investigate the crystal growth of pure W or W-rich solid solutions after the crystal nucleation, regardless the time scale of nucleation. The preferred c-W orientation (011) (the most close packed bcc plane) is consistent with our experiments as confirmed by X-ray diffraction (XRD) in the Bragg-Brentano configuration (not shown; below we show results of XRD in the grazing incidence configuration). The second template of 2000 atoms of a-W not only allowed us to study the growth of these potentially crystalline films before or during the crystal nucleation, but especially facilitates the analysis: difference of results obtained on both templates is a strong fingerprint of a (nano)crystal formation on the former one. The growth simulations of W-poor MGs are possible on both templates (after rapid amorphization on the former) and confirm each other. On both templates, the effect of [W], energy of arriving atoms ($E$ = 1 to 500 eV) and deposition angle ($\theta$ = 0 to 60°) are investigated. The role of $E$ and $\theta$ is examined at two different [W] values of 42% (amorphous composition) and 82% (crystalline composition). The studies of the effect of [W] from 0 to 100% (i.e., both the simulations and the experiments included for comparative purposes also pure W prepared in the same way) were carried out at $E$ = 1 eV and $\theta$ = 0°.

*Contact author: hataali@ntis.zcu.cz

The templates were divided into two regions: the first is a frozen or fixed region located at the bottom, which has been set to a temperature of 0 K to prevent any drifting due to momentum transfer from incident atoms, and the second is an unfrozen or isothermal region in the rest of the template at a temperature of 300 K (NVT ensemble; Nose-Hoover thermostat) that absorbs the kinetic energy of deposited atoms. On each growth template, 3000 atoms were deposited during 750 steps of the recursive algorithm (4 atoms in each step), applying periodic boundary conditions in horizontal directions to approximate an infinite plane. The film atoms deposited were fully relaxed in a NVE ensemble (at fixed energy) during a 1.5 ps long MD run to reach the equilibrium state (this duration is sufficient to cool the thermal spike). The Verlet algorithm with a time step of 0.0005 ps was used in all MD simulations, and the thermostat damping time was 0.01 ps. The incident angle was controlled by the angle between the negative x and z axis of velocities: $v_x$=-$v_0$ cos($\theta$); $v_z$=-$v_0$ sin($\theta$); $v_y$=0, where $v_0$=$\sqrt{2E/m}$ is the initial velocity calculated using mass $m$ of Cu, Zr or W and energy of arriving atoms $E$. Finally, to minimize the statistical noise and improve the reliability, all simulations were performed 5 times and the results were averaged.

### B. Experiment

The results of growth simulations are compared with experimental results of non-reactive magnetron sputtering of Cu-Zr-W films in 0.533 Pa of Ar, also performed at [Cu]/[Zr] ratio close to 1 and [W] varied from 0 to 100%. The films were deposited onto unheated, unbiased, rotating Si(100) substrates in an AJA International ATC 2200-V system using three unbalanced 2" magnetrons: dc sputtering of Zr and W targets and high power impulse sputtering (HiPIMS) of a Cu target. The elemental composition was controlled by sputtering powers (up to 250 W) in the case of Zr and W and deposition-averaged sputtering power (via frequency, 5-19 Hz at a voltage pulse length of 200 µs) in the case of HiPIMS of Cu. The films were subsequently characterized using the analytical techniques described in our previous work dealing with W-free Cu-Zr [31].

## III. RESULTS AND DISCUSSION

### A. Crystallinity and densification

The film structure is illustrated by the final snapshots from eleven growth simulations, as shown in Fig. 1. For the crystalline template c-W(011) (Fig. 1(a)), it can be observed that the Cu-Zr-W films formed an amorphous structure at low [W], and that the structure changes from amorphous to crystalline when more Cu and Zr atoms are replaced by W. For the amorphous template a-W (Fig. 1(b)), the films exhibit an amorphous structure in the whole [W] range, indicating that the nucleation of bcc W takes place at a longer time scale and/or due to higher energies of at least some of the incident atoms. The visual inspection of film structures obtained by simulations on c-W(011) is supported by both experimental and simulated XRD patterns, shown in Fig. 2. In the experimental data (Fig. 2(a)), broad diffraction peaks are observed for films with [W] up to 67%, indicating an amorphous structure. Beyond this threshold, sharp peaks emerge, and new small peaks appear, signifying a structural transition to the bcc structure. In the simulation results in Fig. 2(b), the onset of crystallinity is observed at a slightly lower [W], probably due to a different substrate used for experiments (Si) and simulations (c-W(011); a reliable interaction potential is available for Cu-Zr-W but not for Cu-Zr-W-Si) and possibly also due to the size effect (3.7×3.5 nm horizontal cell size). The dominant reflection of W in the films represents the aforementioned (110) lattice plane and the corresponding peak positions in the simulated XRD patterns (38.75° to 40.05°) are close to those in the measured ones (38.15° to 40.25°) (PDF Card No. 00–004–0806) [46].

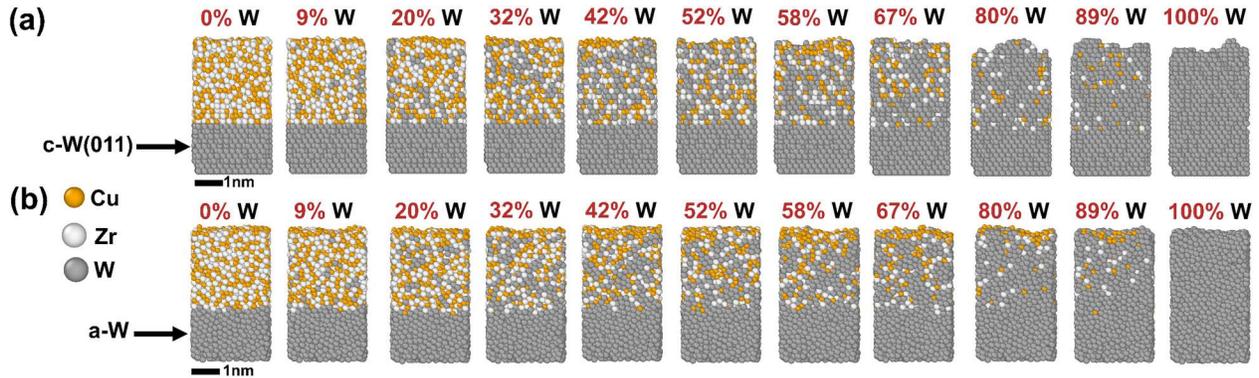

FIG 1. Snapshots of simulated deposited morphologies at eleven selected [W] values, incident energy $E = 1$ eV and deposition angle $\theta = 0°$. Panels (a) and (b) correspond to crystalline the c-W(011) growth template (3.7×3.5×2.1 nm; 1920 atoms) and amorphous a-W growth template (3.6×3.6×2.6 nm; 2000 atoms), respectively, with arrows denoting the top of each template. The balls represent different atom types: Cu (small, dark orange), Zr (large, bright white), and W (medium-sized, dark gray). A total of 3000 atoms are deposited on both growth templates.

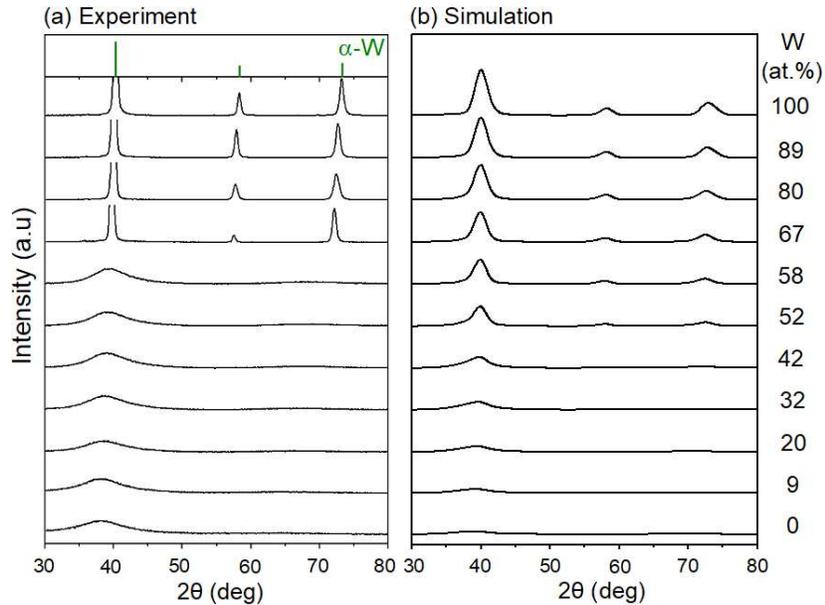

FIG 2. X-ray intensity vs. 2θ taken from Cu-Zr-W films deposited at varied [W] in two different studies: a) Experiment and b) MD simulations. The main diffraction peaks of bcc α-W are marked. The MD simulation results are presented for the c-W(011) template case.

*Contact author: hataali@ntis.zcu.cz

The densification of films deposited can be examined in terms of packing factor ($f_p$), calculated as the ratio of the volume of atoms in the film to the total volume of the film, using atomic radii based on lattice constants of unary metals which the EAM interaction potential leads to ($d_{Cu}$ = 2.55 Å, $d_{Zr}$ = 3.22 Å and $d_W$ = 2.74 Å). Fig. 3 shows the dependence of $f_p$ on the incorporation of W.

For the c-W(011) template, it can be seen that in the purely amorphous range the densification of films decreases with increasing [W] from 0 to 42%. A case can be made that on the one hand, the significant size difference between Cu and Zr allows the Cu atoms to fill holes between Zr atoms, leading to high $f_p$ (even higher than $f_p$ = 74% of pure fcc Cu and hcp Zr). On the other hand, the partial replacement of Cu and Zr by W apparently leads to a combination of atomic diameters which does not fit together equally well, and makes the densification also kinetically difficult due to high mass ($m_W$). This behavior is opposite to previously observed [9] increasing $f_p$ of Cu-Zr MGs after a partial replacement of Zr by Al, which is (i) larger than W (closer to the halfway between Cu and Zr) and (ii) significantly lighter than W. In the [W] range from 42 to 58%, $f_p$ slightly increases (and at [W] = 67% still remains relatively high), possibly as a fingerprint of ordering and formation of first nanocrystals (below the XRD limit). This explanation is supported by the fact that this $f_p$ increase is observed only on the c-W(011) template which facilitates the nanocrystal formation, not on a-W. The dominance of W atoms at [W] > 67% leads to a formation of larger crystals, but also to a formation of defects in these crystals, both vacancies and larger pores (clusters of vacancies). This is related to short diffusion distances of W adatoms during the thermal spikes (note both the high diffusion energy 0.98 eV of W/W(011) [47] and once again high $m_W$) and indicates that the structures obtained are largely given by the growth process rather than just nucleation. On the a-W template, the densification of the films monotonically decreases with increasing [W] in the whole compositional range, for the same reason as that of amorphous films grown on c-W(011). When [W] exceeds 80%, the $f_p$ resulting from the nucleation process on a-W even becomes slightly higher than that resulting from the aforementioned defect-generating growth process on c-W(011).

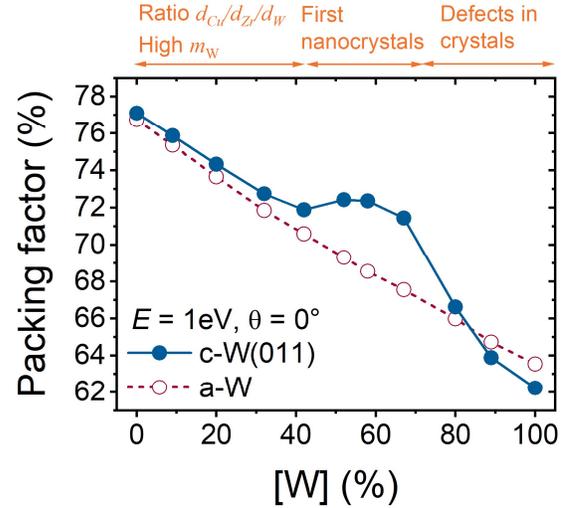

FIG 3. The average packing factor of Cu-Zr-W thin films grown by MD simulations as a function of W content at incident energy $E$ = 1eV and deposition angle $\theta$ = 0° for c-W(011) template (full circles & solid line) and a-W template (empty circles & dashed line). On the former template, the behavior has three distinct regions: i) Suboptimum $d_{Cu}/d_{Zr}/d_W$ ratio together with kinetics limited by high $m_W$, ii) Formation of defect-free (nano)crystals and iii) Defect formation in larger crystals.

**B. Short-range order**

The fundamental concept for good understanding of the structure of Cu-Zr-W thin films is based on studying short-range order (SRO) and medium-range order (MRO) of atoms. First, SRO identifies the first coordination shell of an atom and bonds created with its nearest neighbors. Second, MRO results from the bonding of nearest neighbors of an atom (based on SRO), allowing one to study features involving more than two atoms (network rings and common neighbors).

The first part of the SRO is focused on studying the bonding structure of Cu-Zr-W thin films. This analysis is based on calculating radial distribution functions for each pair of elements, considering that atoms in the first peaks are bonded. Fig. 4 shows the dependence of the fraction of Cu-Cu, Cu-Zr, Cu-W, Zr-Zr, Zr-W, and W-W bonds on [W]. The results show similar bonding statistics on both templates: the formation of W-based (nano)crystals on c-W(011) does not, e.g., lead to a higher fraction of W-W bonds at a cost of Cu-W and Zr-W bonds. This indicates that other atoms are incorporated in the nanocrystals, with consequences, e.g., in terms of compressive stress (see below). At [W] = 0%, the fractions of Cu-Cu and Cu-Zr bonds are significantly lower and slightly higher, respectively, compared to what would correspond to a random placement of atoms into a network. This can be explained not only by higher diameter and in turn coordination of Zr compared to Cu (see next), but also by the fact that the weak Cu-Cu bonds are easier to break during thermal spikes than stronger Zr-Zr bonds (see also the cohesive energies of Cu and Zr in the Supplemental Materials [42] including Refs. [43–45]). The fractions of all W-free bonds monotonically decrease with increasing incorporation of W atoms, which promotes the formation of Cu-W, Zr-W, and W-W bonds. The higher coordination of Zr contributes not only to the aforementioned domination of Zr-Zr bonds over Cu-Cu bonds, but also to that of Zr-W bonds (up to 27%) over Cu-W bonds (up to 21%). In the case of nanocrystal formation on c-W(011) template, the fractions of Zr-W and Cu-W bonds exhibit concave dependencies with maxima at different [W] values of 42% and 58%, respectively, indicating an easier incorporation of small light Cu atoms into the nanocrystals.

*Contact author: hataali@ntis.zcu.cz

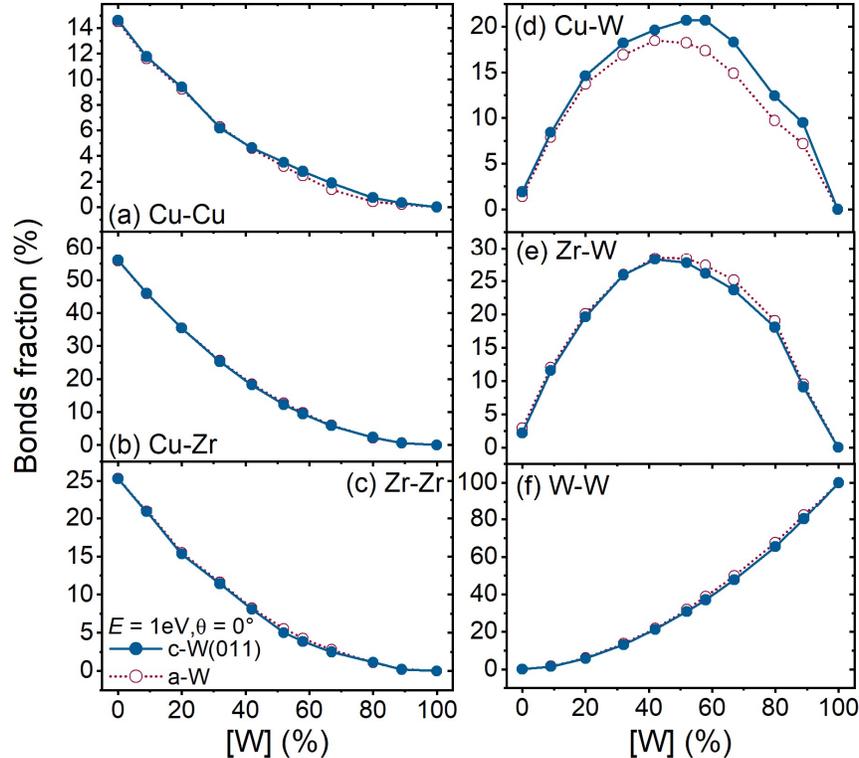

FIG 4. The bonding structure of Cu-Zr-W thin films (number of bonds of each type as a fraction of the total number of bonds, including a small amount of bonds resulting from intermixing with the W template) grown by MD simulations as a function of W content at incident energy $E = 1$eV and deposition angle $\theta = 0°$ for c-W(011) template (full circles & solid line) and a-W template (empty circles & dashed line). Panels (a), (b), (c), (d), (e), and (f) represent Cu-Cu, Cu-Zr, Cu-W, Zr-Zr, Zr-W, and W-W bonds, respectively.

The second part of the SRO examines the average coordination number for each element, shown in Fig. 5. While the coordination of Zr atoms is almost independent of the growth template, the coordination of Cu and W atoms is enhanced due to the (nano)crystal formation on the c-W(011) template. This is in agreement with the observation based on bonding statistics (Figs. 4(d) and 4(e)) that the light Cu atoms (slightly smaller than W) are more easily incorporated into the W based (nano)crystals than the heavier Zr atoms (significantly larger than W). The coordination numbers are calculated from the number of nearest neighbors (in the case of bcc W this includes not only the nearest 8 atoms but also the 6 only slightly more distant atoms), which is affected by the atomic size. Cu is the smallest atom in the films and in turn exhibits the lowest coordination of around or even below 11 (amorphous films), increasing to ≈12 after crystallization on the c-W(011) template. Medium-sized W atoms exhibit intermediate coordination of ≈13 (amorphous films) and up to ≈13.5 (approaching 14 in perfect bcc W) after crystallization on the c-W(011) template. Conversely, the largest atom, Zr, exhibits the highest coordination of ≈14, slowly decreasing with [W] independently of the crystallinity and consistently with the decreasing packing factor (Fig. 3).

*Contact author: hataali@ntis.zcu.cz

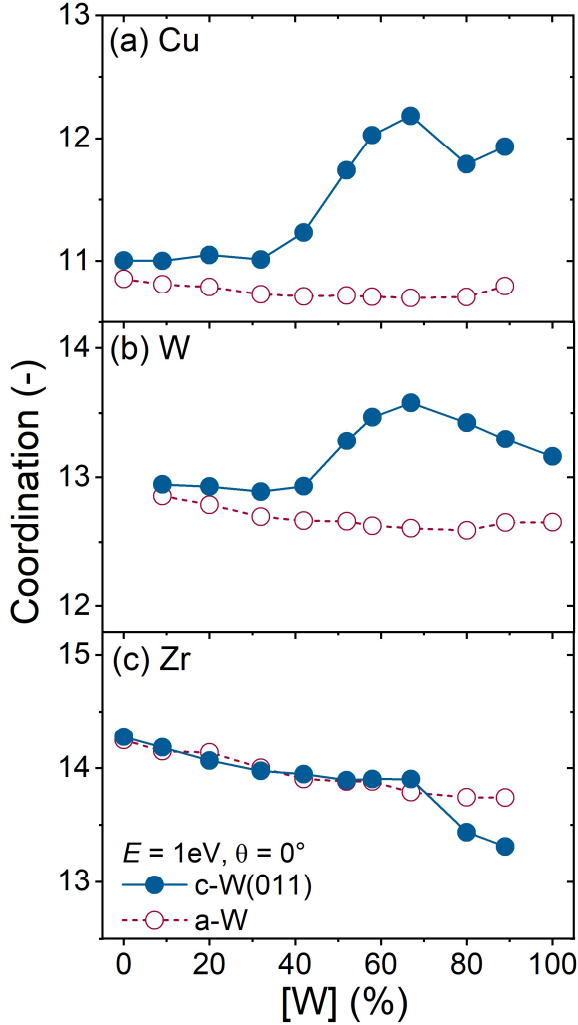

FIG 5. Average coordination numbers of individual elements in Cu-Zr-W films grown by MD simulations as a function of W content, at incident energy $E = 1$ eV and deposition angle $\theta = 0°$ for c-W(011) template (full circles & solid line) and a-W template (empty circles & dashed line).

### C. Medium-range order

The second aspect of the structural study investigates the network topology of Cu-Zr-W thin films. This analysis is quantified using shortest-path (SP) network ring statistics and common neighbor analysis. In the context of SP network ring statistics [48], the crystal quality is quantified in terms of rings which have no shortcuts, i.e. contains the shortest path between two atoms. The present study considers three distinct crystalline structures: fcc, hcp, and bcc, which correspond to preferred unit cells of Cu, Zr, and W, respectively. The fcc and hcp structures are distinguished by the presence of 8 3-membered rings and 3 4-membered rings per atom, while the bcc structure is characterized by 12 3-membered rings and 3 4-membered rings per atom. In the context of metallic glasses, the icosahedral clusters, which represent the primary structural unit of MGs exhibit 50 3-membered rings and 12 5-membered rings per 13-atom cluster.

Fig. 6 shows the dependence of the network ring statistics of Cu-Zr-W thin films on [W]. For the amorphous structures, [W] does not have a convincing effect on the MRO. However, for the growth on the c-W(011) template, all statistics exhibit fingerprints of the transition from MG through bcc nanocrystal formation to larger defect-containing bcc crystals, let us recall the visual inspection (Fig. 1(a)) and X-ray intensity (Fig. 2). Most of the changes take place when [W] increases from 32 to 67% (consistently with $f_p$ in Fig. 3). Fig. 6(a) shows increasing number of 3-membered rings (present in both icosahedral clusters and bcc crystals, but there are more of them per atom in the latter), Fig. 6(b) shows increasing number of 4-membered rings (only in bcc) and Fig. 6(c) shows decreasing number of 5-membered rings (present only in icosahedral clusters). Above [W] = 67% the crystal quality in terms of 3- and 4-membered rings does not further rise, in agreement with the aforementioned defect formation.

*Contact author: hataali@ntis.zcu.cz

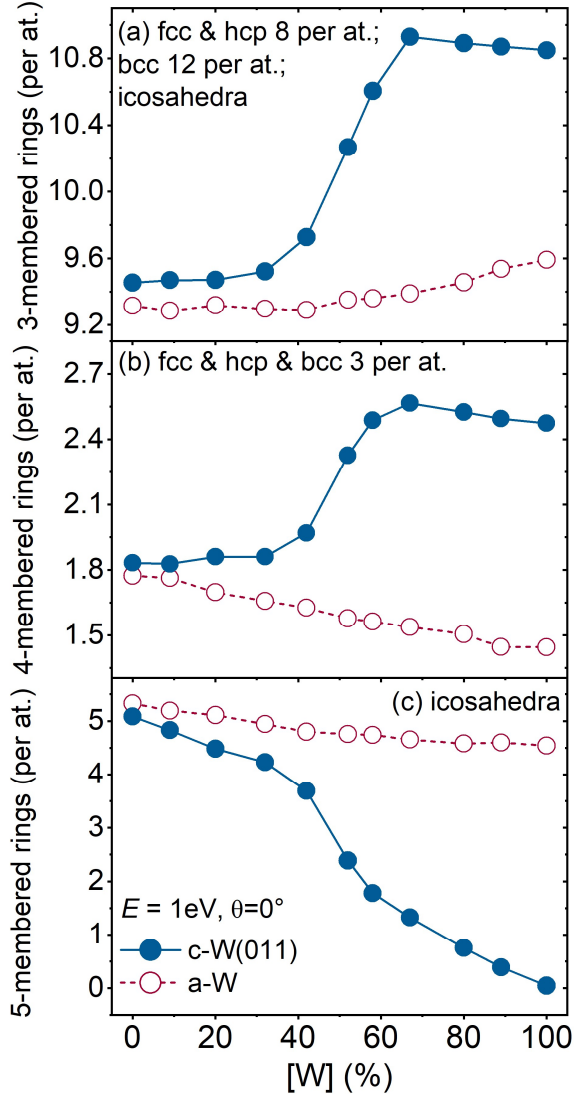

FIG 6. The shortest path network ring statistics for the Cu-Zr-W films grown by MD simulations as a function of W content, at incident energy $E$ = 1eV and deposition angle $\theta$ = 0° for c-W(011) template (full circles & solid line) and a-W template (empty circles & dashed line). Panels (a), (b), and (c) represent the number of shortest-path network rings (per film bulk atom) of the length 3, 4, and 5, respectively (the occurrence of longer rings is negligible).

In parallel to the SP network rings, we express the MRO in terms of Common Neighbor Analysis (CNA) which is based on counting the number of *ijk* triplets where *i* is the number of common neighbors of two bonded atoms, *j* is the number of bonds between two of these neighbors, and *k* is the length of the longest chain which can be formed using these bonds [49].

Let us recall that the Cu atoms crystallize in an fcc structure, corresponding to 6 triplets 421 per atom. Zr atoms crystallize in a hcp structure with 3 triplets 421

*Contact author: hataali@ntis.zcu.cz

+ 3 triplets 422 per atom, and W atoms crystallize in a bcc structure, featuring 4 triplets 666 + 3 triplets 444 per atom. Finally, there are 12 triplets 555 per 13-atom icosahedral cluster. These triplets are valid for perfect crystals and clusters, while in real materials, many triplets can change due to a single defect (example for 555 below).

Fig. 7(a) shows all CNA triplets in Cu-Zr-W thin films at [W] = 42% (predominantly amorphous composition, but at the edge of the range of nanocrystal formation). It can be seen that the most dominant triplets in the films grown on both templates are icosahedral-like triplets 433, 544 and 555 (the former triplets come from the perfect icosahedral cluster e.g. by breaking one surface bond, changing 12×555 to 8×555 + 2×544 + 2×433 per central atom). However, the visually amorphous films grown at [W] = 42% on c-W(011) exhibit also a considerable concentration of bcc-like triplets 444 and 666. This confirms the nanocrystal formation (below the X-ray detection limit), even more convincingly than the other quantities such as packing factor or numbers of network rings.

Figs. 7(b)-(f) show how do the concentrations of selected triplets 411, 421, 555, 666 and 444 (crystals and perfect clusters) depend on [W]. Again, for the growth on the c-W(011) template, the analysis exhibits fingerprints of the transition from MG through bcc nanocrystal formation to larger defect-containing bcc crystals. First, there is a direct confirmation that, indeed, no other crystalline structure than bcc forms. The concentrations of fcc- and hcp-like triplets 421 and 422 (Fig. 7(b) and 7(c), respectively), rather low already in the Cu-Zr MGs at [W] = 0%, further decrease toward zero after W incorporation. Second, the icosahedral clusters (decreasing concentration of triplets 555 in Fig. 7(d)) are gradually replaced by the bcc (nano)crystals (increasing concentration of triplets 666 and 444 in Fig. 7(e) and 7(f), respectively). Yet again (like in the case of $f_p$ and network rings), most of the changes take place between [W] = 32 and 67%.

Note that the central atom of perfect icosahedral clusters must be surrounded by 12 atoms ($N$ = 12). The average coordination numbers in the amorphous compositions ($N_{Cu} \approx 11$, $N_W \approx 13$ and $N_{Zr} \approx 14$; Fig. 5) make Cu and W much more likely candidates than Zr for occupying the central positions of these clusters. Detailed analysis (not shown graphically) revealed that the icosahedral clusters are centered mostly around Cu at [W] = 0-20%, almost equally around Cu and W at [W] = 32% and mostly around W at [W] = 42-58% (followed by a comparatively negligible cluster concentration at even higher [W]).

In the amorphous films grown on a-W, the number of triplets 421, 422, 444 and 666 is negligible compared to the number of icosahedral-like triplets

555. The central atoms of icosahedral clusters with $N = 12$ are Cu (at low [W]) or W (at high [W]) also here. However, Figs. 3 and 7(d) collectively reveal one of the very specific consequences of W incorporation into $Cu_xZr_xW_{100-2x}$ MGs: increasingly icosahedral-like medium-range order in parallel to decreasing packing factor. This is contrary to, e.g., incorporation of Al into $Cu_{46}Zr_xAl_{54-x}$ MGs [9], where the different combination of atomic radii led to increasingly icosahedral-like medium-range order in parallel to increasing packing factor.

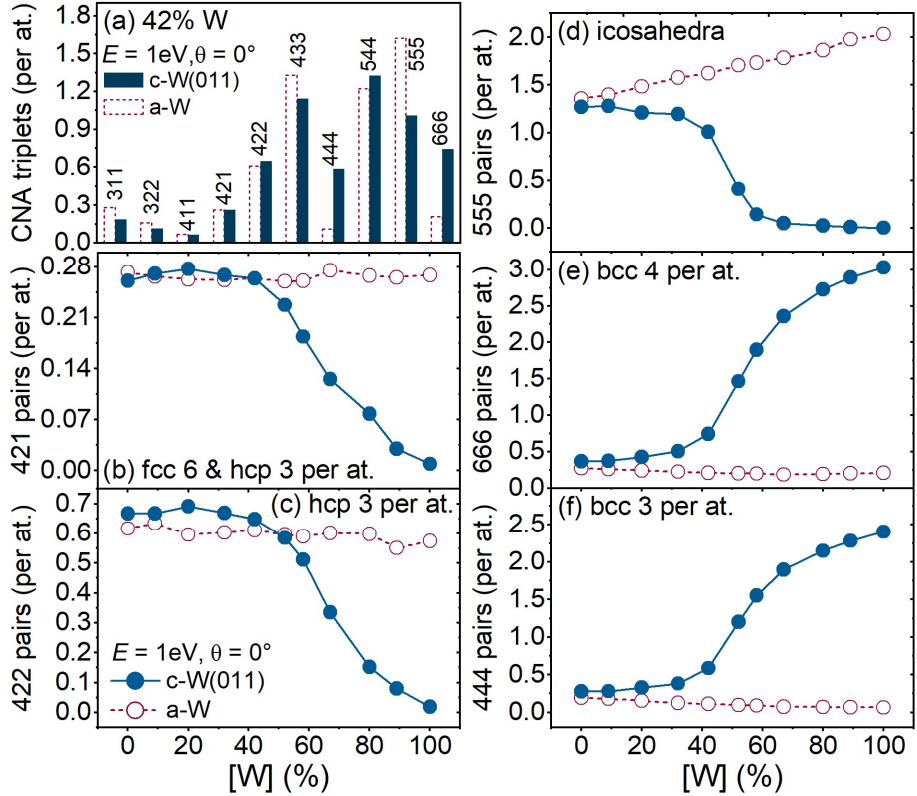

FIG 7. Common Neighbor Analysis (CNA) for the Cu-Zr-W films grown by MD simulations as a function of W content, at incident energy $E = 1$ eV and deposition angle $\theta = 0°$ for c-W(011) template (full columns or full circles & solid line) and a-W template (empty dashed columns or empty circles & dashed line). Panel (a) shows the histogram of all CNA triplets at [W] = 42%. Panels (b), (c), (d), (e), and (f) represent the number of selected CNA triplets (per film bulk atom) 421, 422, 555, 666, and 444, respectively.

## D. Mechanical properties

The study of the mechanical properties of materials is crucial for improving their performance and ensuring their safety in various application fields. These properties are strongly influenced by the structure of the materials. Therefore, it is essential to analyze the factors controlling the structural formation (see above) not only in order to understand the structural formation itself, but also in order to understand their impact on the mechanical properties of the materials. In this context we aim to analyze the stress in Cu-Zr-W thin films, given by averaging per-atom stress tensors (including the contributions of atomic kinetic energy and interatomic forces). The stress equation is explained in detail in previous works [50,51].

Fig. 8 shows the stress evolution of Cu-Zr-W thin films as a function of [W]. For the amorphous structures, [W] does not have a convincing effect on the (very slightly tensile) stress. However, for the growth on the c-W(011) template, the gradual formation of densified (nano)crystals at [W] ≤ 67% generates compressive stress down to -1.5 GPa. The stress has to be expected owing to the fact that the bcc W-based crystals contain also other atoms (in the first place Cu, as discussed above), and it is consistent with the local maximum of packing factor shown in Fig. 3. At [W] > 67%, the stress is relaxed via the formation of vacancies and other even larger defects (decreasing packing factor in Fig. 3).

*Contact author: hataali@ntis.zcu.cz

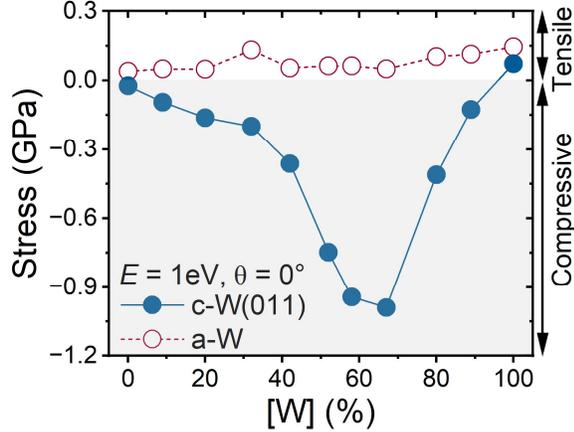

FIG 8. The calculated stress in Cu-Zr-W thin films grown by MD simulations as a function of W content, at incident energy $E = 1$eV and deposition angle $\theta = 0°$ for c-W(011) template (full circles & solid line) and a-W template (empty circles & dashed line). Compressive stress has negative values (grey area) and tensile stress has positive values (white area).

The film properties obtained by simulations, particularly the densification, MRO and stress obtained on the c-W(011) template, are complemented by those obtained experimentally and shown in Fig. 9. The measured properties are collectively controlled by the structure and by the character of covalent/metallic bonding in the materials. The covalent component of the bonding, crucial for the mechanical properties, is particularly strong for W and particularly weak for Cu (let us recall the corresponding melting temperatures and the cohesive energies in the Supplemental Materials [42] including Refs. [43–45]). However, owing to the structural evolution, none of the measured mechanical properties monotonically increases along the replacement of Cu and Zr by W. Instead, the measured dependencies exhibit the most interesting features in the same narrow compositional range as the calculated dependencies.

The significant hardness increase from 5.4 to 15.0 GPa (Fig. 9(a)) takes place between [W] = 0 and 67%, i.e. up to the exactly the same limit which has been identified by simulations as an onset of the replacement of densified nanocrystals by larger defect-containing crystals and of the stress relaxation. The hardness increase is arguably due to the (i) stronger interatomic bonds and (ii) stress, while it is not correlated with the packing factor. At [W] > 67% the hardness approximately saturates, indicating that the possible factors increasing the hardness (stronger interatomic bonds, optimum volume fractions of the crystalline and the amorphous component in a nanocomposite) and decreasing the hardness (stress relaxation, defect formation, replacement of a nanocomposite by a single phase in the case of pure W) approximately compensate each other.

The other mechanical properties measured, hardness to effective Young's modulus ratio (measure of the elastic strain to failure $H/E_{eff}$ in Fig. 9(b)) and elastic recovery (Fig. 9(c)), exhibit features even more far from a monotonic increase with [W]. There are concave dependencies on [W], with maxima achieved in both cases at [W] = 52%, i.e. in the middle of the compositional range characterized by significant formation of densified nanocrystals and local maximum of the packing factor. While the W incorporation has been successful in terms of the enhancement of $H/E_{eff}$ from 0.055 to 0.098 and of the elastic recovery from 35 to 61%, both these quantities decrease at too high [W].

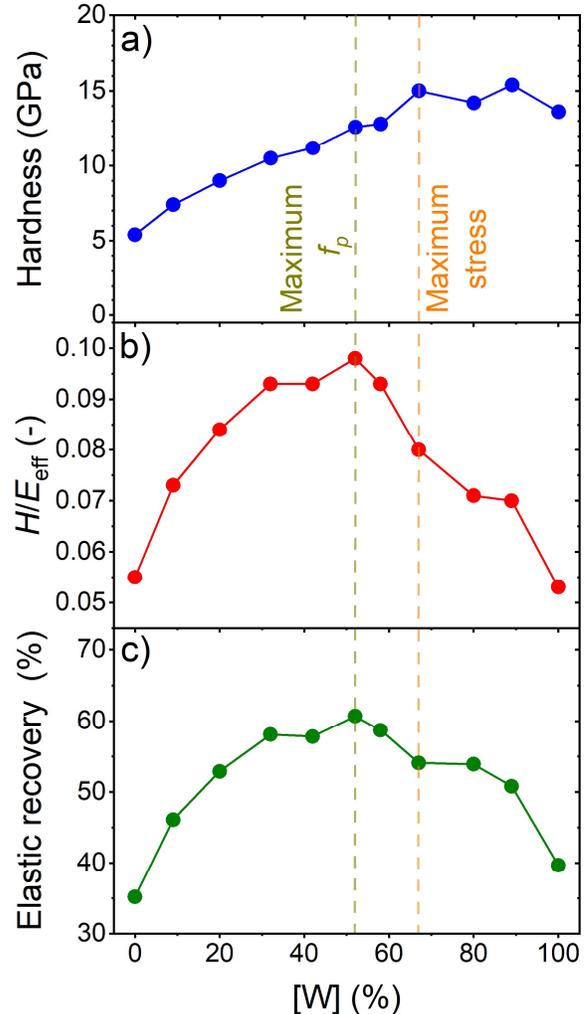

FIG 9. Measured properties of Cu-Zr-W films deposited at varied [W] by magnetron sputtering. Panel (a) shows the hardness, panel (b) shows the ratio of hardness and Young's modulus and panel (c) shows the elastic recovery.

*Contact author: hataali@ntis.zcu.cz

**E. Effect of growth conditions**

To present the selected most interesting effects of the incident energy and angle on film characteristics, two different [W] values are considered: [W] = 42% (visually amorphous composition) and [W] = 80% (crystalline composition when using the c-W(011) template). Fig. 10 shows the effect of energy on the stress in Cu-Zr-W thin films. While the low value $E$ = 1 eV which is used above is relevant for many experiments, the figure predicts the role of accelerated metal ions (especially in the case of high degree of ionization in the case of HiPIMS) and the role of high-energy tails of energy distribution functions of both metal neutrals and ions. It can be seen that the stress behavior is qualitatively the same for both compositions and also for both growth templates. At low energies from 1 to 60-100 eV, the stress in the films exhibits a convex evolution with a minimum (i.e. most compressive) value at $E_{threshold}$ = 20 ± 10 eV. The visual inspection (not shown) revealed that at $E$ < $E_{threshold}$ the incoming atoms can break the bonds between the atoms at the top of the islands formed during the initial deposition phase. These three-dimensional islands then transform directly into two-dimensional clusters which coalesce to form dense films. As the energy increases above $E_{threshold}$, the kinetic energy of arriving atoms is too high to be easily dissipated, and the atoms diffuse and mix with the atoms at the interface between the film and the substrate. This process can lead to the formation of defects and voids within the film structure, and the stress becomes less compressive on c-W(011) and more tensile on a-W. The convex stress-energy dependence is consistent with experimental results of amorphous covalent materials [52], captured by the subplantation model proposed by Davis [53] and Robertson [54]. At very high $E$ above 100 eV, the film stress gets more compressive once again for both compositions and templates since the number of substrate atoms diffusing towards the film-substrate interface increases, thereby increasing the diffusion of atoms within the thin films (this last result is of course relevant mostly for the initial growth stage). As a result, the former defects and voids are filled, resulting in a dense and compact structure.

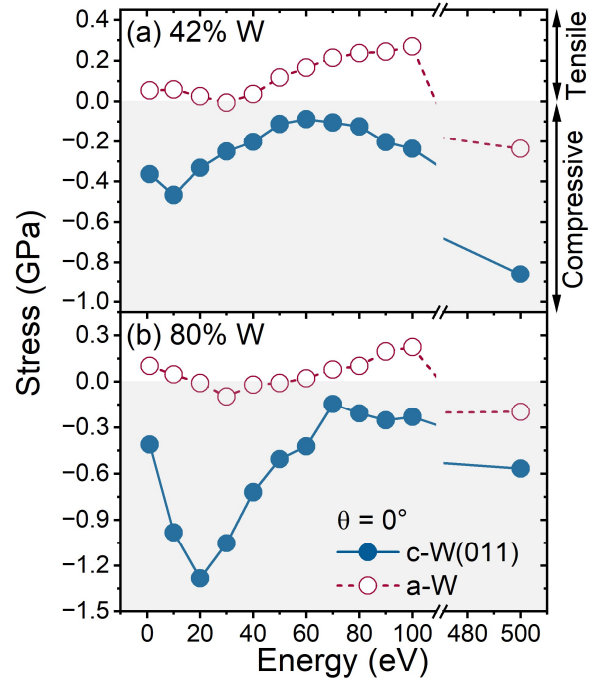

FIG 10. The calculated stress in Cu-Zr-W thin films grown by MD simulations as a function of incident energy at a deposition angle θ = 0° and [W] = 42% (visually amorphous; panel (a)) or 80% W (crystalline on c-W(011); panel (b)) for c-W(011) template (full circles & solid line) and a-W template (empty circles & dashed line). Compressive stress has negative values (grey area) and tensile stress has positive values (white area).

The deposition angle θ is an important factor in producing thin films with different structures, such as nanorods, helices and chevrons [55], which can be used in many applications. In this context, the films have been grown by simulations using not only a film-forming flux perpendicular to the (XY) plane (θ = 0°) but also an inclined film-forming flux (θ ≠ 0°).

Fig. 11 shows the simulated stress in Cu-Zr-W films as a function of deposition angle, once again for both aforementioned [W] values of 42% and 80% and both growth templates. On the one hand, the figure shows that the difference between stresses obtained on c-W(011) (compressive due the gradual formation of densified nanocrystals) and on a-W (slightly tensile), observed above at θ = 0°, is observed also at other θ values below the threshold for nanorod formation (θ ≤ 40° for [W] = 42% in Fig. 11(a) and θ ≤ 20° for [W] = 80% in Fig. 11(b)). Furthermore, $f_p$ is approximately constant, meaning the structures are still dense in this range (see Fig. S1 in Supplemental Materials [42]).

However, when the angle is increased above this threshold, the stress obtained on c-W(011) changes toward or even above zero (that is, the stress becomes less dependent on the growth template). $f_p$ decrease due to the change in structure from films to nanorods on both templates, as discussed in Fig. S1. More information on the nanorods formation process can be found in Fig. S2 in Supplemental Materials [42]. The nanorods formed exhibit attractive forces at their boundaries, resulting in a tensile stress. Also these explanations are in agreement with experimental studies [56].

*Contact author: hataali@ntis.zcu.cz

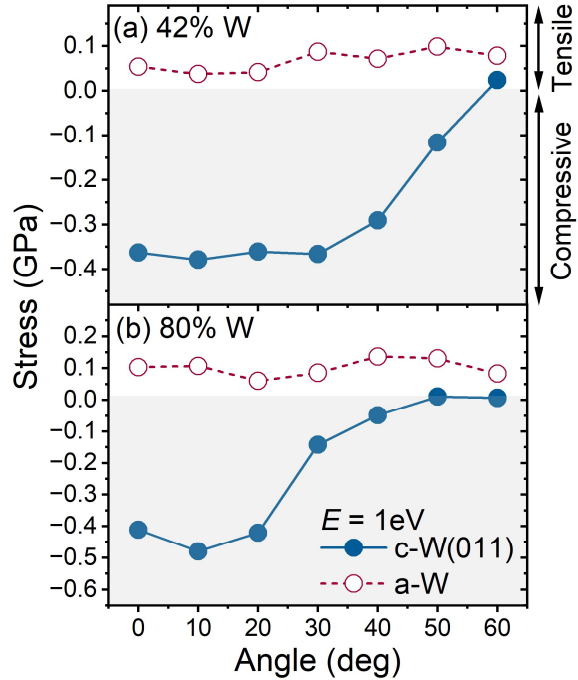

FIG 11. The calculated stress in Cu-Zr-W thin films grown by MD simulations as a function of deposition angle at an incident energy $E = 1eV$ and [W] = 42% (visually amorphous; panel (a)) or 80% W (crystalline on c-W(011); panel (b)) for c-W(011) template (full circles & solid line) and a-W template (empty circles & dashed line). Compressive stress has negative values (grey area) and tensile stress has positive values (white area).

## V. CONCLUSIONS

Molecular dynamics simulations and magnetron sputtering have been used to study the effect of W in thin films of metallic glasses based on $Cu_{50}Zr_{50}$ in a wide range of compositions and growth conditions. The experimental XRD analysis, specifically the transition from amorphous ([W] = 0 to 58%) to crystalline ([W] = 67 to 100%) films, is consistent with the visual inspection of films grown by simulations on crystalline template. However, detailed analysis of the packing factor and medium-range order in terms of network ring and common neighbor statistics revealed that these films are truly amorphous only between [W] = 0 and 32%. At [W] = 42-52% the films are visually amorphous but possess fingerprints of dense nanocrystals (including a dependence of film characteristics on the growth template, i.e. on the conditions for nanocrystal formation, and leading to a local maximum of the packing factor). At [W] = 58-67% the films are visually crystalline, at preserved densification. At [W] = 80-100% there is an increasing concentration of vacancies and even larger defects in the crystals (and a decreasing packing factor regardless the crystallinity), related to the short diffusion distances of heavy W. The analysis of the short-range order reveals presence of other atoms (in the first place small light Cu) in the W-based (nano)crystals, and supports the explanation of a maximum compressive stress at [W] = 58-67% when the crystals are already large, but the stress is not yet relieved by the vacancy formation. The results obtained on amorphous films further demonstrate the importance of which element is incorporated into Cu-Zr: e.g. the relationship between the packing factor and the icosahedral-like medium-range order in Cu-Zr-W is the opposite than in Cu-Zr-Al. The experimental characterization of the effect of W on materials' properties includes an improvement of hardness from 5.4 to 15.0 GPa, hardness to effective Young's modulus ratio from 0.055 to 0.098 and elastic recovery from 35 to 61%. The specific dependencies on [W] are in an excellent agreement with and are explained by the growth simulations: the improvements take place in the same range as the main structural changes, while the hardness at the best saturates at [W] > 67% (onset of the defect formation and stress relaxation in the simulations) and the other two quantities even decrease at [W] > 52% (local maximum of the packing factor in the simulations). In addition to the aforementioned simulations of the low-energy normal angle growth, further simulations allowed us to identify the energies of incident atoms leading to a maximum compressive stress ($E = 20 \pm 10$ eV depending on the composition and crystallinity) and three threshold angles leading to a conversion of almost dense films to nanorods (θ > 20-50° also depending on the composition and crystallinity). Collectively, the results are important for understanding the role of W in modifying the structure and improving the mechanical performance of Cu-Zr metallic glasses, predict optimum compositions which maximize some of the mechanical properties, and contribute to the development of advanced materials for various applications.

## ACKNOWLEDGMENTS
This work was supported by the Czech Science Foundation under Project No. GA22-18760S and by the project QM4ST under Project No.

*Contact author: hataali@ntis.zcu.cz

CZ.02.01.01/00/22_008/0004572 funded by Program Johannes Amos Comenius, call Excellent Research. Computational resources were provided by the e-INFRA CZ project (ID:90254), supported by the Ministry of Education, Youth and Sports of the Czech Republic.

*Contact author: hataali@ntis.zcu.cz

*Contact author: hataali@ntis.zcu.cz